# Mesons and Resonances in Relativistic Quantum Mechanics For the Lorentz-Scalar Potential

Mikhail Nicolas Sergeenko

Stepanov Institute of Physics, the National Academy of Sciences of Belarus, Minsk, Belarus
msergeen@usa.com

**Abstract:** Mesons as bound states of quark and anti-quark in the framework of a relativistic potential model are studied. Interaction of constituents in bound state is described by the Lorentz-scalar QCD inspired funnel-type potential with the coordinate dependent strong coupling, $\alpha_S(r)$. Lagrangian relativistic mechanics is used to derive the main dynamic two particle equation of motion. On this basis, relativistic two body wave equation is derived. Solution of the equation for the system in the form of a standing wave is obtained. Two exact asymptotic expressions for the meson squared mass are obtained and used to derive the meson universal mass formula. Light and heavy meson mass spectra are calculated.

## 1. Introduction

Hadron data represent the purest imprint of the hadronic world. The number of known hadrons is constantly increasing with the growing energies of accelerators. Thus, in the last decade more than ten new charmonium-like states have been discovered. More and more states will be discovered in the near future. According to the Particle Data Group [1], many hadrons are still absent from the summary tables. Obviously, there is still a lot of work to be done both theoretically and experimentally.

Most particles listed in the PDG tables are unstable. A thorough understanding of the physics summarized by the tables is related to the concept of a resonance. Resonance is a widely known phenomenon in Nature and our everyday life. This is the tendency of a system to oscillate at a greater amplitude at some frequencies, which are known as the system's resonant frequencies. Resonance may be defined as the excitation of a system by matching the frequency of an applied force to a characteristic frequency of the system. Because the frequencies resonate, or are in sync with one another, maximum energy transfer is possible.

The properties of some states are still not very clear. There are theoretical indications that some of these new states could be the first manifestation of the existence of exotic hadrons (tetraquarks, molecules, hybrids etc.), which are expected to exist in Quantum Chromo Dynamics (QCD) [2].

A vast amount of experimental data on hadron spectroscopy raises the question of their classification and establishing mass relations. The eightfold way and the standard SU(3) Gell-Mann-Okubo (GMO) formula [3,4] have played an important role in the historical progress in particle physics. However, the direct generalization of the GMO formula to the charmed and bottom hadrons cannot agree well with experimental data due to higher-order breaking effects. There are many works focused on the mass relations, including inequalities [5,6] and equalities [7,8].

A unique pre-QCD approach in hadron physics represents the Regge theory [9], which is based on general analyticity and crossing properties of scattering amplitudes. All mesons and baryons in this theory are associated with Regge poles which move in the complex angular momentum plane as a function of energy. The Regge trajectory of a particular pole, the total angular momentum $J = \text{Re}\{\alpha(M^2)\}$ (resonance effective spins), is characterized by a set of internal quantum numbers. Sets of multiple Regge trajectories can be reduced to a single line, representing an entire meson family ($\pi$, $\rho$, $\varphi$, etc.).

In the Regge theory, hadrons are classified in terms of the Regge trajectories. The GMO classification scheme works for light hadrons which correspond to the linear Regge trajectories. However, it was shown that the Regge trajectories are complex nonlinear functions, which can be described with the use of the potential models [10,11]. This means that the Regge trajectories can be considered as an alternate classification method.

At the present time, there is no strict theory of mesons and resonances as relativistic two-body systems. The description of the systems as quark-antiquark bound states and their excitations in a way fully consistent with all requirements imposed by special relativity and within the framework of relativistic Quantum Field Theory (QFT) is one of the great challenges in theoretical elementary particle physics. Within the framework of QFT the covariant description of relativistic bound states is the Bethe-Salpeter (BS) formalism [12].

The description of hadronic properties which strongly emphasizes the role of the minimum-quark-content part of the wave function of a hadron is generically called the quark model. It exists on many levels: from the simple, almost dynamics-free picture of strongly interacting particles as bound states of quarks and antiquarks, to more detailed descriptions of dynamics, either through models or directly from QCD itself.

Semirelativistic potential models have been proved extremely successful for the description of mesons. The main characteristics of the meson spectra can be obtained with a spinless Salpeter (SS) equation [13,14] supplemented with the Cornell interaction (a Coulomb-like potential plus a linear confinement) [15]. Numerous techniques have been developed in order to solve numerically with a great accuracy the semirelativistic equation. Nevertheless, it is always interesting to work with analytical results. Several attempts to obtain some mass formulae for hadrons were already performed. Some approaches rely on fundamental QCD properties [16], but they are limited to the study of ground states of hadrons. In other works, the hadron masses are given by a completely phenomenological point of view [17]. We will adopt here a point of view by assuming that a semirelativistic potential model allows a good description of the main features of meson spectra.

In this work we study mesons as relativistic two body systems in the potential approach. The problems encountered are related to 1) two-particle relativistic equation of motion and 2) absence of a strict definition of the potential in relativistic theory. The Coulomb potential is treated as the Lorentz-scalar function of the spatial variable $r$. The concept of position dependent particle mass is developed. We use an assumption that in relativistic kinematics the spatial two particle relative momentum is relativistic invariant, derive a relativistic two-particle wave equation and obtain its asymptotic solution in the form of a standing wave. The free particle hypothesis for the bound state is developed: quark and antiquark move as free particles inside of the meson. The relative motion of quarks in eigen states is described by the standing wave of the form $C_n \sin(k_n x + \delta_n)$ for each spatial degree of freedom.

## 2. The potential in relativistic theory

The potential is correctly defined in nonrelativistic (NR) theory. The NR potential model has proven extremely successful for the description quark-antiquark bound states. The use of the NR potential in relativistic kinematics gives even more accurate results for meson spectra. This success is somewhat puzzling in that it persists even when the model is applied to relativistic systems. Potential models work much better than one would naively expect [11,14–16]. Why the NR potential model is so successful even in totally relativistic theories like Regge method in high energy hadron physics [9]?

The potential in relativistic theory has not been as rigorously defined as it is taken to be in classical mechanics. In relativistic mechanics one faces with different kind of speculations around the potential, because of absence of a strict definition of the potential in this theory. In many applications an object's "potential energy"' is a more useful dynamical quantity than is the "force" being exerted on the object. In fact, for the quantum world the concept of force does not exist and the potential energy function replaces it as the prime quantity of interest. How does one use the potential energy function to deduce motion in classical physics? If a particle's potential energy graph and its total energy are known, then all dynamical quantities of the particle can be found.

### 2.1. The vector-like Coulomb potential

The interaction of a relativistic particle with the four-momentum $p_\mu$ moving in the *external* field $A_\mu$ is introduced according to the gauge invariance principle, $p_\mu \to P_\mu = p_\mu - A_\mu$. For zero component $P_0 = E/c$ this results in the total energy of the particle in the field of the form (here we use the system, in which $\hbar = c = 1$)

$$E = \sqrt{\mathbf{p}^2 + m^2} + V(r), \qquad (1)$$

where the potential $V(r) = eA_0(r)$ is vector-like, i.e., additive to the four-component of $p_\mu$. This equation by the correspondence principle (replacing physical quantities by operators acting onto the wave function) gives the one-particle SS equation [13,14]. However, in case of a bound state, we have a closed system, no external field and any particle of the system can be considered as moving source of the interaction field. In this case the interacting particles and the potential is a unit system.

The Coulomb potential is of fundamental importance in all physics. It is used in relativistic theory as well as in the NR potential models. A good test of potential models is given by their application to the description of atomic spectra. The binding energy of an electron in a static Coulomb field (the external electric field of a point nucleus of charge $Ze$ with infinite mass) can be determined in NR formulation from solution of the Schredinger's equation. The NR description of the atomic spectra is accurate enough; however, it is not exact. The usual approach is to take the nonrelativistic approximation as the starting point. Then corrections are applied using perturbation theory.

A serious problem of relativistic potential models is definition and the nature of the potential: whether it is Lorentz-vector or Lorentz-scalar or their mixture [18]? The spectroscopic data are usually analyzed with the use of the Somerfield's Fine-Structure Formula [19],

$$E_{nj} = m_e c^2 \left\{ 1 + \left[ \frac{Z\alpha}{n - (j + \tfrac{1}{2}) + \lambda(j)} \right]^2 \right\}^{-1/2}, \qquad (2)$$

where $\lambda(j) = [(j + \tfrac{1}{2})^2 - (Z\alpha)^2]^{1/2}$, $n = 1, 2, \ldots$ is principal and $j$ is total angular momentum quantum numbers. The energy levels of atoms are determined mainly by the eigenvalues (2), QED effects such as self energy and vacuum polarization, nuclear size and motion effects. The eigenvalues (2) as the ones predicted from solution of the Klein-Gordon wave equation, obtained for the Lorentz-vector Coulomb potential [18,20,21].

One should note the following about the eigenenergies (2). The term $\lambda(j)$ becomes complex if $Z\alpha > j + \tfrac{1}{2}$. This means that the $S$ states start to be destroyed above $Z = 137$, and that the $P$ states begin destroyed above $Z = 274$. Note that this differs from the result of the Klein-Gordon equation, which predicts $S$ states being destroyed above $Z = 68$ and $P$ states destroyed above $Z = 82$. Besides, the radial $S$-wave function $R(r)$ diverges as $r \to 0$, i.e., $R(r) \approx r^\beta$, with $\beta = -\tfrac{1}{2} + \lambda(0)$, and there are no normalizable solutions for vector-like potentials [22]. What is the reason of the problems? This problem is very important in hadron physics where, for the vector-like confining potential, there are no normalizable solutions.

### 2.2. The scalar-like Coulomb potential

A Schrodinger-like relativistic wave equation of motion for the Lorentz-scalar position-dependent potential $S(r)$ was formulated in [20,21]. Though the physical meaning of the Lorentz-scalar potential is not well described, it is generally accepted that the Lorentz-scalar potential is coupled to the rest mass of the particle such that the relativistic energy-momentum relations are given as [10,20,21]

$$E^2 - \mathbf{p}^2 = \mathsf{m}^2(r), \quad (3)$$

where $\mathsf{m}(r) = m_0 + S(r)$ is the position dependent scalar mass. This relation is a consequence of the Lagrange equations of relativistic motion, with the chosen relativistic Lagrangian and a scaled time as the evolution parameter. This issue was investigated in our previous works [10,11,15,23] for quarkonia and gluebolls [24,25]. It was shown that the effective interaction has to be Lorentz-scalar in order to confine quarks and gluons.

A particle of mass $m$ and charge $-e$ moving around a nucleus of charge $+Ze$ under an interaction of the Lorentz- scalar Coulomb potential was considered [20,21]. Solution of the time-independent Schrodinger-like relativistic wave equation for the potential in polar spherical coordinates was obtained. The predictions of this formalism are free from the Klein's paradox, and for the Lorentz-scalar Coulomb potential, this formalism yields the exact bound-state eigenfunctions and eigenenergies [21],

$$E_{nj} = mc^2 \left\{ 1 - \left[ \frac{Z\alpha}{n - (j + \frac{1}{2}) + \lambda'(j)} \right]^2 \right\}, \quad (4)$$

where $\lambda'(j) = [(j + \frac{1}{2})^2 + (Z\alpha)^2]^{1/2}$. The energy eigenvalues (4) are the same as those predicted exactly from solution of the relativistic semi-classical wave equation [23].

Application of this Schrödinger-like formalism for the Lorentz-scalar square-step potential was shown to be free from the Klein's paradox [20,21]. The predictions are free from not only the Klein's paradox, but also the paradoxical results predicted for the Lorentz-vector potential mentioned above [22].

Thus, in general there are two different relativistic versions: the potential is considered either as the zero component of a four-vector, or as a Lorentz-scalar. The relativistic correction for the case of the Lorentz-vector potential is different from that for the case of the Lorentz-scalar potential. A comparison between these two cases was given in [21,22]. In contrast to the Lorentz-vector potential, the radial $S$-wave function $R(r)$ for the Lorentz-scalar one is regular as $r \to 0$, i.e., $R(r) \sim r^\beta$ with $\beta = -\frac{1}{2} + \lambda'(0)$ [21]. Also, there are normalizable solutions for scalar-like potentials [20–23]. The Lorentz-scalar potential results in the concept of position dependent mass (3), as usually investigated [20,21,23].

### 2.3. The Cornell potential

There is some unity in the world of meson spectroscopy. The universal flavour-independent confining potential is fixed in an extremely simple manner in terms of very small number of parameters, all of which have a direct physical interpretation. It is generally agreed that, in perturbative QCD (pQCD), as in Quantum Electro Dynamics (QED) the essential interaction at small distances is instantaneous coulomb one-gluon exchange (OGE); in QCD, it is $qq$, $qg$, or $gg$ Coulomb scattering. Therefore, one expects from OGE a Coulomb-like contribution to the potential at $r \to 0$,

$$V_\mathrm{S}(r) \Box - \frac{\alpha_\mathrm{S}}{r}, \quad (5)$$

where $\alpha_\mathrm{S}$ is the strong coupling constant. For large distances, in order to be able to describe confinement, the potential has to rise to infinity. The long range part is dominated by a confinement that lattice calculations predict linear in the quark distance, i.e., $V_\mathrm{L}(r) \sim \sigma r$ for large $r$, where the string tension is $\sigma \approx 0.15$ GeV$^2$. The short-distance coulomb term, $V_\mathrm{S}(r)$, and long-distance linear one, $V_\mathrm{L}(r)$, by simple summation, $V(r) = V_\mathrm{S}(r) + V_\mathrm{L}(r)$, lead to the famous funnel-shaped quark-antiquark (Cornell) potential [15],

$$V_{q\bar{q}}(r) = -\frac{4}{3}\frac{\alpha_S}{r} + \sigma r. \quad (6)$$

The Cornell potential (6) is a special in hadron physics. It incorporates the basic physical quantities of the strong interaction: one-gluon exchange at small and the string tension at large distances. All phenomenologically acceptable QCD-inspired potentials are only variations around the Cornell potential (6). This potential is unique in that sense, if considered in the complex-mass scheme; it yields the complex eigenmasses for hadrons and resonances, predicting their masses and widths [11].

### 2.4. The modified Cornell potential

The Cornell potential (6) is fixed in an extremely simple manner in terms of very small number of parameters, which are directly related to basic physical quantities noted above. However, the strong coupling $\alpha_\mathrm{S}$ in (6) is a free parameter. This potential can be modified by introducing the $\alpha_\mathrm{S}(r)$ dependence [24].

The QCD running coupling in momentum representation is a function of the virtuality $Q^2$, $\alpha_\mathrm{S}(Q^2 = |q^2|)$, or $\alpha_\mathrm{S}(r)$ in the configuration space. More accurate calculations of hadronic masses require the accounting for the dependence $\alpha_\mathrm{S}(r)$ in the potential (6). To find the dependence $\alpha_\mathrm{S}(r)$, we use the concept of dynamically generated gluon mass, $\mu(Q)$, which arises from an analysis of the gluon Dyson-Schwinger (DS) equations [26].

The infinite set of couple DS equations cannot be resolved analytically. An approximate resolution of the DS equations was obtained by Cornwall in the Feynman gauge. In Euclidean space ($Q^2 > 0$), the strong coupling is [26]

$$\alpha_\mathrm{S}(Q) \equiv \frac{g^2(Q)}{4\pi} = \frac{1}{b_0 \ln\left\{\left[Q^2 + 4\mu^2(Q)\right]/\Lambda^2\right\}}, \quad (7)$$

where $b_0 = (33 - 2n_\mathrm{f})/(12\pi)$, $n_\mathrm{f}$ is number of flavors, $\Lambda$ is the QCD scale parameter. The nonperturbative (NP) QCD generates an effective, momentum-dependent mass, without affecting the local SU$_c$(3) invariance, which remains intact. This solution contains a dynamically generated gluon mass $\mu(Q)$ and is another NP approach which has led to a very appealing physical picture. The gluon mass generation is a purely NP effect associated with the existence of infrared finite solutions for the gluon propagator.

The use of the mnemonic rule, $Q \to 1/r$, in (7) results in the ansatz [24],

$$\alpha_S(r) \equiv \frac{g^2(r)}{4\pi} = \frac{1}{b_0 \ln\left[(1+4\mu_g^2 r^2)/\Lambda^2 r^2\right]}, \quad (8)$$

for the strong running coupling in the coordinate space. The running coupling (8) conserves the basic properties of (7) in the momentum representation: it agrees with the asymptotic freedom properties, i.e., $\alpha_S(r) \to 0$ at $r \to 0$. Because of the presence of the dynamical gluon mass the strong effective charge, $g(Q)$, extracted from these solutions freezes at a finite value, giving rise to an infrared fixed point for QCD,

$$\bar{\alpha} \equiv \alpha_S(r \to \infty) = \left[2b_0 \ln(2\mu_g/\Lambda)\right]^{-1}, \quad (9)$$

where $\mu_g = \mu(Q)$ is the gluon mass at $Q \to 0$, and $\mu(Q) \to 0$ at $Q \to \infty$ ($r \to 0$). Thus, with the help of (9), one comes to the following potential of interaction [25]:

$$V_{q\bar{q}}(r) = -\frac{4}{3}\frac{\alpha_S(r)}{r} + \sigma r, \quad (10)$$

The spin-dependent corrections to the potential (10) can also be included [10]. The modified funnel-type potential (10) is considered here to be Lorentz-scalar in order to confine quarks and gluons inside hadrons.

### 2.5. Relativistic properties of the Coulomb potential

Consider some relativistic properties of the Coulomb potential. This potential as a function in 3$D$-space is given by the propagator $D(\mathbf{k}^2) = -1/\mathbf{k}^2$ (Green function) in momentum space. The Fourier transform of the function $4\pi\alpha D(\mathbf{k}^2)$ gives the Coulomb potential $V(r) = -\alpha/r$. In Minkowski 4$D$-space, a free photon propagates as massless particle with the four-momentum $k_\mu$ which is on-mass shell, i.e., $k^2 = k_\mu k^\mu = k_0^2 - \mathbf{k}^2 = 0$. But, in case of the potential the photons are virtual, the interaction is carried by the virtual photons and described by the propagator, $D(k^2) = 1/k^2$. In this case, the virtual photon (as a carrier of interaction) is off-mass shell, i.e. $k_0^2 - \mathbf{k}^2 \neq 0$.

The Coulomb potential in the four-momentum space [27],

$$\tilde{V}(t) = 4\pi\frac{\alpha}{t}, \quad (11)$$

where $t = k^2 < 0$ is momentum transfer, corresponds to the scattering amplitude in the Born approximation with the photon's propagator $D(k^2)$. The Fourier transform of (11) is

$$U(t,\mathbf{r}) = 8\pi^2\alpha\int\frac{d^4k}{(2\pi)^4}\frac{-1}{k^2}e^{-ikx} = -\frac{\alpha}{|r|}\left[\delta(ct-r) - \delta(ct+r)\right], \quad (12)$$

where $x = (t, \mathbf{r})$, leads to the Coulomb potential.

The squared momentum transfer, $k^2 = k_0^2 - \mathbf{k}^2 < 0$. This means that the virtual photon's invariant mass, $\mu = \pm i\sqrt{k^2}$, is imaginary. Similar situation we observe in Regge theory of strong interactions, where reggeons (which are described by the Regge trajectories) are hypotetical particles with imaginary mass.

These arguments could be explanation of the instantaneous Coulomb interaction, because the imaginary-mass particle propagates with the velocity $v > c$. The $\delta$-functions in (12) mean two directions in space-time, i.e. the "forward-backward" exchange by the virtual photon. The arguments $ct \pm r = 0$ of the $\delta$-functions mean that the photon's proper time, $\tau = 0$, the relative distance $r = |\mathbf{r}|$ is Lorentz invariant, therefore, the interaction is instantaneous. Thus, two expressions (11) and (12) could be considered as a definition of the Lorentz-invariant Coulomb potential.

## 3. Relativistic two body equations

There are several forms of relativistic two-particle wave equations such as the Klein-Gordon equation, the Dirac equation, quasipotential equations [28,29]. The homogeneous BS equation governs all the bound states. Attempts to apply the BS formalism to relativistic bound-state problems give series of difficulties such as the impossibility to determine the BS interaction kernel beyond the tight limits of perturbation theory [30]. Its inherent complexity usually prevents to find the exact solutions or results in the appearance of excitations in the relative time variable of the bound-state constituents (abnormal solutions), which are difficult to interpret in the framework of quantum physics [30].

For various practical reasons and applications to both QED and QCD some simplified equations, situated along a path of NR reduction, are used. More valuable are methods which provide either exact or approximate analytic solutions for various forms of differential equations. They may be remedied in three-dimensional reductions of the BS equation; the most well-known of the resulting bound-state equations is the one proposed by Salpeter [13,14].

### 3.1. The spinless Salpeter equation

There exist various reductions of the two-body BS equation [28,29]. Under suitable simplifying assumptions, one can derive analytically examples of rigorous solutions to the instantaneous homogeneous BS equation by relating tentative solutions to the interactions responsible for formation of bound states [14].

In the BS approach a bound state of four-momentum $P_\mu$ and squared mass $M^2 = P_\mu P^\mu$ is described by a BS amplitude defined in configuration-space representation as matrix element of the time-ordered product of the field operators of the bound-state constituents. In momentum space representation the BS amplitude encodes the distribution of the relative momentum $p$ of the two bound-state constituents.

The amplitude satisfies the formally exact homogeneous BS equation, which involves two kinds of dynamical ingredients: the full propagators $S_i(p_i)$ of the constituents with individual momenta $p_i$ ($i = 1, 2$), and the BS interaction kernel $K(p_1, p_2, P)$, by construction a fully amputated four-point Green function of the bound-state constituents defined perturbatively by summation of the countable infinity of all BS-irreducible Feynman diagrams for two particles into two-particle scattering.

The most straightforward way out is the reduction of the BS equation to the SS equation by a series of approximations, which rely on some simplifying assumptions [14]. 1) Eliminate any dependence on time like variables; the instantaneous approximation assumes that in the center-of-momentum frame (c.m.f.) of the bound states, fixed by $P_\mu = (M, \mathbf{P} = 0)$, all interactions between bound state constituents are *instantaneous*. In this ("static") limit, the BS interaction kernel $K(p_1, p_2, P)$ depends only on the spatial com-ponents, $p_1$ and $p_2$, of the relative 4-momenta $p_1$, and $p_2$ involved, that is, takes the form $K(p_1, p_2, P) = K(\mathbf{p}_1, \mathbf{p}_2)$. 2) Assume that the BS kernel $K(\mathbf{p}_1, \mathbf{p}_2)$ depends only on the difference $\mathbf{q} = \mathbf{p}_1 - \mathbf{p}_2$, $K(\mathbf{p}_1, \mathbf{p}_2) = K(\mathbf{q})$ which means that $K(\mathbf{q})$ is of convolution type, which entails for the potential $V(\mathbf{p}, \mathbf{q}) = V[(\mathbf{p}-\mathbf{q})^2]$; trivially, this means that each momentum-space potential $V(\mathbf{p}, \mathbf{q})$ is the Fourier transform of a spherically symmetric configuration-space potential $V(r)$. 3) Neglect any reference to the spin degrees of freedom of the involved bound-state constituents. 4) Restrict the whole formalism exclusively to positive-energy solutions, which will be denoted by wave function $\psi(\mathbf{r})$. 5) Any bound-state constituent is assumed to propagate as *free* particle with some effective mass $m_i$ required to encompass appropriately all dynamical self-energy effects [14].

The reduction of the BS equation to the SS equation in [30] relies on exactly two simplifying assumptions: 1) the instantaneous $(t = t_1 - t_2 = 0)$ approximation assumes that in the c.m.f. of the bound states, fixed by $P_\mu = (\mathsf{M}, \mathbf{P} = \mathbf{0})$, all interactions between bound-state constituents are instantaneous. In this limit, the BS interaction kernel $K(p,q,P)$ depends only on the spatial components, $\mathbf{p}$ and $\mathbf{q}$, of the relative four-momenta $p$ and $q$ involved takes the form $K(p, q, P) = K(\mathbf{p},\mathbf{q})$. 2) Any bound-state constituent is assumed to propagate as *free* particle with some effective mass $m_i$ required to encompass appropriately all dynamical self-energy effects.

After applying all these simplifying assumptions and approximations to the BS equation, one comes to the SS equation in the c.m.f. [13,14],

$$\hat{H}\psi(\mathbf{r}) = \mathsf{M}\,\psi(\mathbf{r}),  \quad (13a)$$

$$\hat{H} = \sqrt{(-i\nabla)^2 + m_1^2} + \sqrt{(-i\nabla)^2 + m_2^2} + V(r). \quad (13b)$$

This is the simplest relativistic two-particle eigenvalue wave equation. Here in (13a) $\mathsf{M}$ is the mass of the system, the potential $V(r)$ arises as the Fourier transform of the BS kernel $K(\mathbf{q})$. It is sometimes denoted semirelativistic since is not a covariant formulation. However, even this simplest two-particle wave equation (13) leads to difficulties. The square root of the operators cannot be used as it stands; it would have to be expanded in a power series before the momentum operator, raised to a power in each term, could act on $\psi(\mathbf{r})$. As a result of the power series, the space and time derivatives are completely asymmetric: infinite-order in space derivatives but only first order in the time derivative, which is inelegant and unwieldy.

The next problem is the noninvariance of the energy operator in (13), equated to the square root which is also not invariant. Another more severe problem is that it can be shown to be nonlocal and can even violate causality: if the particle is initially localized at a point $\mathbf{r}_0$ so that $\psi(\mathbf{r}_0, t = 0)$ is finite and zero elsewhere, then at any later time $t$ the equation predicts delocalization $\psi(\mathbf{r}_0, t = 0) \neq 0$ everywhere, even for $r > ct$ which means the particle could arrive at a point before a pulse of light could. This would have to be remedied by the additional constraint $\psi(\mathbf{r}_0, t = 0) = 0$.

There is another problem: the $S$-wave solution of the equation (13) for the Coulomb potential diverges at the spatial origin and behaves as $\psi(r) \sim r^{-4\alpha/3\pi}$ at $r \to 0$ [22]. This divergence at the spatial origin is actually a general problem affecting relativistic wave equations with the Lorentz-vector potential. For example, the solution of the Dirac equation with the Coulomb potential for the $S$-wave states behaves as $\psi(r) \sim r^\lambda$, $\lambda = (1 - \alpha^2)^{1/2} - 1$ at $r \to 0$ [31].

### 3.2. Relativistic Quantum Mechanics (RQM)

Another approach to bound-state problem which is close to nonrelativistic one is RQM. The formulation of RQM differs from NR Quantum Mechanics (QM) by the replacement of invariance under Galilean transformations with invariance under Poincare transformations. The RQM is also known in the literature as relativistic Hamiltonian dynamics or Poincare-invariant QM with direct interaction.

This theory is applicable to massive particles propagating at all velocities up to those comparable to the speed of light $c$, and can accommodate massless particles. The theory has application in high energy physics, particle physics and accelerator physics, as well as atomic physics, chemistry and condensed matter physics. RQM is only an approximation to a fully self-consistent relativistic theory of known particle interactions because it does not describe cases where the number of particles changes.

The dynamics of many-particle system in the RQM is specified by expressing ten generators of the Poincare group $M^{\mu\nu}$ and $P_\mu$ in terms of dynamical variables. In the constructing generators for interacting systems it is customary to start with the generators of the corresponding noninteracting system; the interaction is added in the way that is consistent with Poincare algebra. In the relativistic case it is necessary to add an interaction $V$ to more than one generator in order to satisfy the commutation relations of the Poincare algebra.

There are three equivalent forms in the RQM called "instant", "point", and "light-front" forms [31]. The description in the "instant" form implies that the operators of three-momentum and angular momentum do not depend on interactions. The description in the "point" form implies that the mass operators $M^{\mu\nu}$ are the same as for noninteracting particles and these interaction terms can be presented only in the form of the four-momentum operators $P_\mu$. There is no unique way of separating the generators into dynamical subset (including interaction $V$) and kinematical subset, which must be associated with some subgroup of Poincare group, usually called stability group or kinematics subgroup [31].

The construction for two interacting particles proceeds as follows. The two-particle Hilbert space of noninteracting system is defined as tensor product of two one-particle Hilbert spaces. A two-body unitary representation of Poincare group on the two-particle Hilbert space is reducible. A basis in this space can be constructed from single-particle bases [31].

Clebsch-Gordan coefficients for Poincare group are constructed and used to reduce the unitary representation on the two-particle Hilbert space to linear superposition (direct integral) of irreducible representations. Poincare generators for irreducible representations of the noninteracting system are constructed, along with operators for the mass, spin, helisities.

# 4. The two body problem in RQM

The SS equation (13) is the conceptually simplest bound-state wave equation incorporating to some extent relativistic effects. This equation is obtained with the help of several simplifying assumptions; it has to be regarded as a well-defined standard approximation to the BS formalism. However, (13) can be obtained by another way without any approximations in the framework of RQM from the following simple and obvious consideration.

The total energy $E$ of the system in NR classic mechanics is given by the sum of the *kinetic* $T(p)$ and the *potential* $V(r)$ energies, $E = T(p) + V(r)$. The total energy $E$ is the integral of motion. Similar approach can be used in relativistic theory and the problem under investigation can be considered as follows. Two free scalar particles 1 and 2 are characterized by their coordinates $q_1$, $q_2$ in Minkowsky spacetime and the four-momenta $p_1$, $p_2$. If the particles interact each other and create a system, we introduce the additional four-vector $Q^\mu(q_1, q_2)$, some function describing the interaction. This four-vector is *not* an external field, but is a part of the system. The two particles and the vector-function $Q^\mu(q_1, q_2)$ together represent a closed system, which can be described by the four-momentum,

$$P^\mu = P_0^\mu(p_1, p_2) + Q^\mu(q_1, q_2), \tag{14}$$

where the $P_0(p_1, p_2) = p_1 + p_2$. The square of the four momentum (14), $P_\mu P^\mu = \mathsf{M}^2$, is the system's squared invariant mass.

The four-vector (14) describes free motion of the bound-state system and can be separated into two equations,

$$E = \sqrt{\mathbf{p}_1^2 + m_1^2} + \sqrt{\mathbf{p}_2^2 + m_2^2} + Q^0(q_1, q_2) = \text{const}, \tag{15}$$

$$\mathbf{P} = \mathbf{p}_1 + \mathbf{p}_2 + \mathbf{Q}(q_1, q_2) = \text{const}. \tag{16}$$

The equations (15) and (16) describe the energy and momentum conservation laws of the bound state system, i.e., the total energy $E$ and the total momentum $\mathbf{P}$ are the constants of motion. If the bound system is stable, the Hamiltonian (15) can not depend on the time explicitly, the function $Q(q_1, q_2)$ should not depend on time [32], but depends on the positions $\mathbf{r}_1$ and $\mathbf{r}_2$ of particles, i.e., $Q(q_1, q_2) = Q(\mathbf{r}_1, \mathbf{r}_2)$.

It was discussed in Section 3.1 that in the c.m.f. of the bound states, fixed by $P_\mu = (\mathsf{M}, \mathbf{P} = \mathbf{0})$, $Q^\mu = (V, \mathbf{Q} = \mathbf{0})$, all interactions between bound state constituents are instantaneous. In this case, the interaction potential respects spherical symmetry, i.e., $V(\mathbf{q}^2)$, where $\mathbf{q} = \mathbf{p}_1 - \mathbf{p}_2$, is of convolution type.

Coordinates $q_1$, $q_2$ and momenta $p_1$, $p_2$ are conjugate variables, therefore, the relative momentum $\mathbf{q}$ is conjugate to the relative distance $\mathbf{r} = \mathbf{r}_1 - \mathbf{r}_2$. The system's relative time, $\tau = 0$, i.e., the interaction is instantaneous. This means that the Fourier transform of the interaction potential, $V(\mathbf{q}^2)$, defines the interaction $V(r)$ in configuration space. Similar result for the Coulomb potential gives the Fourier transform (12). Thus, (15) and (16) in the c.m.f. take the form

$$\sqrt{\mathbf{p}^2 + m_1^2} + \sqrt{\mathbf{p}^2 + m_2^2} + V(r) = \mathsf{M}, \tag{17}$$

$$\mathbf{p}_1 + \mathbf{p}_2 = \mathbf{0}, \tag{18}$$

where $\mathbf{p} = \mathbf{p}_1 = -\mathbf{p}_2$, is the particle momentum in the c.m.f. The vector $\mathbf{p}$ is the conjugate variable of the inter-distance $\mathbf{r}$.

The equation (17) corresponds to the Hamiltonian (13b) of the SS equation (13a).

The total energy $E$ (mass $\mathsf{M}$ of the system in the c.m.f.) of two free particles with masses $m_1$, $m_2$,

$$E = \sqrt{\mathbf{p}^2 + m_1^2} + \sqrt{\mathbf{p}^2 + m_2^2}, \tag{19}$$

can be transformed to the invariant squared momentum,

$$\mathbf{p}^2 = \frac{1}{4s}\lambda(s, m_1^2, m_2^2) \equiv \frac{s - m_-^2}{4s}(s - m_+^2) = \kappa^2, \tag{20}$$

where $s = \mathsf{M}^2$ is the Mandelstam's invariant, $m_- = m_1 - m_2$, $m_+ = m_1 + m_2$, and $\lambda(m_1, m_2, s)$ is *basic invariant kinematical (three-angle) function* [9]. In case of interacting particles, it is generally accepted that the Lorentz-scalar potential is coupled to the rest mass of particle such that the relativistic energy-momentum relation is given by (3), which is a consequence of the Lagrange equations of relativistic motion [21]. Taking the time derivative of both sides of (3), since the total energy $E$ is constant, one obtains the equation of motion,

$$\gamma^2[m_0 + S(r)]\frac{d\mathbf{v}}{dt} = -\vec{\nabla}S(\mathbf{r}), \tag{21}$$

where $\gamma = (1 - \mathbf{v}^2/c^2)^{-1/2}$. The scaled (proper) time $\tau$ is related to the temporal (coordinate) time $t$ in (21) $dt = \sigma d\tau$, where the scaled factor $\sigma$ is chosen as $\sigma = E/m$, where $E$ is the total energy of the particle. With these notations, the relativistic equation (21) can be written in the form,

$$\frac{d\mathbf{p}}{d\tau} = -\vec{\nabla}S(\mathbf{r}) = \mathbf{F}. \tag{22}$$

This is the Minkowski space-part force acting on the particle. The total relativistic energy and the momentum of the particle are:

$$E = \gamma[m_0 + S(r)], \quad \mathbf{p} = \gamma[m_0 + S(r)]\mathbf{v}. \tag{23}$$

Thus, we accept the invariant Lorentz-scalar distance dependent particle mass.

Similarly, one can consider two relativistic particles with the Lorentz-scalar potential $\mathsf{W}(r)$ of interaction. The relativistic total energy $\varepsilon_i(p)$ of a particle $i$, given by the equation $\varepsilon_i^2(p) = p^2 + m_i^2$, can be represented as sum of the kinetic energy, $T_i(p)$, and the rest mass, $m_i$, i.e., $\varepsilon_i(p) = T_i(p) + m_i$. This is why the system's mass (17) can be written as

$$\mathsf{M} = T_1(\mathbf{p}) + m_1 + T_2(\mathbf{p}) + m_2 + \mathsf{V}(r). \tag{24}$$

Kinetic and potential energies are different types of the total energy. It is natural to combine the Lorentz-scalar potential $\mathsf{W}(r)$ with masses of particles. As its Lorentz structure is not precisely known, we suppose that the confinement is scalar. The scalar potential $\mathsf{V}(r)$ is shared among the two masses $m_1$ and $m_2$; one choice is to take $m_1 + m_2 + \mathsf{V}(r) = [m_1 + \alpha_1 \mathsf{V}(r)] + [m_2 + \alpha_2 \mathsf{V}(r)]$, where $\alpha_1 = m_2/(m_1 + m_2)$, $\alpha_2 = m_1/(m_1 + m_2)$ as in [17]. But, we use other weight coefficients: $\alpha_1 = \alpha_2 = \frac{1}{2}$, so that (24) takes the form:

$$\mathsf{M} = T_1(\mathbf{p}) + \mathsf{m}_1(r) + T_2(\mathbf{p}) + \mathsf{m}_2(r), \tag{25}$$

where we have introduced the position dependent particle masses $\mathsf{m}_i(r) = m_i + \frac{1}{2}\mathsf{V}(r)$, because the potential $\mathsf{V}(r)$, as a part of the bound system, acts equally on each of the particles.

There is another explanation for (25). Lagrangian in RQM is formulated with the proper-time as the evolution parame-ter [20,21]. In case of two relativistic particles with the scalar potential of interaction $V(r)$, the system's proper relative time, $\tau = 0$, (instantaneous interaction), the system's proper time, $T$, as the evolution parameter, is the same for both particles in the bound state. Taking the time derivative of (18) by the proper time $T$ with the help of (22) and (23), we have $F_1 + F_2 = 0$, for the space part of Minkowski force, that gives $|F_1| = |F_2|$. Forces acting on bound particles in the c.m.f. are equal and opposite each other. Therefore, taking into account (22)–(25), one can write, for the system's total energy (invariant mass):

$$M = \sqrt{\mathbf{p}^2 + m_1^2(r)} + \sqrt{\mathbf{p}^2 + m_2^2(r)}. \qquad (26)$$

The equation (26) can be transformed to the form (20) for $\mathbf{p}^2$ as the function of the invariants $s = M^2$, $m_1(r)$ and $m_2(r)$,

$$\mathbf{p}^2 = \frac{s - m_-^2}{4s}\left[s - (m_+ + V)^2\right] \equiv \kappa^2 - U(s,r), \qquad (27)$$

where $\kappa^2$ is the squared invariant momentum given by (20),

$$U(s,r) = \frac{s - m_-^2}{4s}\left[2m_+ V(r) + V^2(r)\right], \qquad (28)$$

is the potential function, $K(s, m_-) = (s - m_-^2)/4s$. The equation (27) is the relativistic analogy of the NR expression $\mathbf{p}^2 = 2\mu[E - V(r)] \equiv k^2 - U(E,r)$, where $\mu = m_1 m_2 / m_+$ for the two body system; it can be written in the form of one-particle equation:

$$M^2 = \boldsymbol{\pi}^2 + [m_+ + V(r)]^2, \qquad \boldsymbol{\pi}^2 \equiv \frac{4s}{s - m_-^2}\mathbf{p}^2. \qquad (29)$$

The constants of motion $M$ and $\kappa^2$ are connected by the relation

$$M = \sqrt{\kappa^2 + m_1^2} + \sqrt{\kappa^2 + m_2^2}. \qquad (30)$$

The expressions (26), (27) and (29) are equivalent and used to write the corresponding two-particle wave equations.

## 5. The two body wave equation and its solution

The wave equations in QM can be derived with the help of the fundamental *correspondence principle*, which has been used at the stage of creation of quantum theory. Equation (26), (27) and (30) by the correspondence principle (replacing physical quantities by operators acting onto the wave function) give three equivalent two-particle spinless wave equations

$$\left[\sqrt{\left(-i\vec{\nabla}\right)^2 + m_1^2(r)} + \sqrt{\left(-i\vec{\nabla}\right)^2 + m_2^2(r)}\right]\psi(\mathbf{r}) = M\psi(\mathbf{r}), \qquad (31)$$

$$\left[\left(-i\vec{\nabla}\right)^2 + U(s,r)\right]\psi(\mathbf{r}) = \kappa^2 \psi(\mathbf{r}), \qquad (32)$$

$$\left\{\frac{4s}{s - m_-^2}\left(-i\vec{\nabla}\right)^2 + [m_+ + V(r)]^2\right\}\psi(\mathbf{r}) = M^2 \psi(\mathbf{r}), \qquad (33)$$

The wave equations (31)–(33) could be solved by regular methods, but not for the potential (10). The equation (32) can be considered as a Schrödinger-type wave equation with relativistic kinematics. It is hard to find an analytic solution of known relativistic wave equations for the potential (10) that does not allow us to get an analytic dependence $E^2(k,l)$. This aim can be achieved with the use of QC method which is the mathematical realization of the correspondence principle in QM. The QC method developed in [33,34] was tested as the exact for *all* one-particle solvable spherically symmetric potentials [33]. The corresponding eigenfunctions have the same behavior as the asymptotes of the exact solutions.

To derive the QC wave equation we start with the corresponding classic equation in Hamilton-Jacobi formulation. In this work, we consider (27); the corresponding wave equation is (32). To obtain the QC equation, we replace the operator $\Delta = \vec{\nabla}^2$ by the *canonical operator* $\Delta^c$. In the spherical coordinates $q = \{r, \theta, \varphi\}$ with the determinant $\det g_{ij} = r^2 \sin\theta$ of the metric tensor, $g_{ij}$, the canonical operator is

$$\Delta^c = \frac{\partial^2}{\partial r^2} + \frac{1}{r^2}\frac{\partial^2}{\partial \theta^2} + \frac{1}{r^2 \sin^2\theta}\frac{\partial^2}{\partial \varphi^2}. \qquad (34)$$

The operator (34) acts in the representation on the state function $\Psi(\mathbf{r}) = \sqrt{\det g_{ij}}\,\psi(\mathbf{r})$. The QC wave equation corresponding to (32) is

$$\left[-\Delta^c + U(s,r)\right]\Psi(\mathbf{r}) = \kappa^2 \Psi(\mathbf{r}). \qquad (35)$$

The normalization condition is given by the equality

$$\int |\Psi(\mathbf{r})|^2 dr d\vartheta d\varphi \equiv \int |\psi(\mathbf{r})|^2 \det g_{\mu\nu} dr d\vartheta d\varphi = 1. \qquad (36)$$

The generalized QC two-particle wave equation in the arbitrary curvilinear coordinates, $Q = \{q_1, q_2, q_3\}$, is

$$\left[\sum_{i=1}^{3}\left(\frac{-i\hbar}{g_{ii}}\frac{\partial}{\partial q}\right)^2 + U(s,q)\right]\Psi(\mathbf{q}) = \kappa^2 \Psi(\mathbf{q}). \qquad (37)$$

It is the second-order differential equation of the Schrödinger type in canonical form. An important feature of this equation is that, for two and more turning-point problems, it can be solved by the conventional leading order in $\hbar$ WKB method [33].

An appropriate solution method of the QC wave equation, which is the same for NR and relativistic systems, was developed in [33,34]. In this method, each of the one-dimensional equations obtained after separation of the QC wave equation is solved by the same QC method. The QC wave equation (35) for the potential (10) is solved by the asymptotic method. This means that (35) is considered separately for two limiting cases, i.e., for the short-distance coulombic, $V_S(r)$, and the long-distance linear part, $V_L(r)$, of the potential (10). Joining the two asymptotic solutions with the help of the two-point Pade approximant we obtain the interpolating mass formula.

The QC equation (35) for the Coulomb-like term of the potential (10) in the spherical coordinates is

$$\left\{-\Delta^c + \frac{s - m_-^2}{4s}\left[s - \left(m_+ - \frac{a(r)}{r}\right)^2\right]\right\}\Psi(\mathbf{r}) = 0. \qquad (38)$$

The equation (38) is separated that gives the radial

$$\left\{\frac{d^2}{dr^2} + \frac{s - m_-^2}{4s}\left[s - \left(m_+ - \frac{a(r)}{r}\right)^2\right] - \frac{M_l^2}{r^2}\right\}R(r) = 0 \qquad (39)$$

and angular,

$$\left(\frac{d^2}{d\vartheta^2} + M^2 - \frac{M_z^2}{\sin^2\vartheta}\right)\Theta(\vartheta) = 0, \quad (40)$$

equations, where $M^2$ and $M_z^2$ are the constants of separation and, at the same time, integrals of motion. The angular QC equation (40) is especially important in our approach since it determines the angular momentum eigenvalues $M_l$; this equation is solved by the same QC method.

The QC quantization condition appropriate to (40) at the interval $0 \le \theta \le \pi$ is [33,34]

$$\int_{\vartheta_1}^{\vartheta_2}\sqrt{M^2 - \frac{M_z^2}{\sin^2\vartheta}}\,d\vartheta = \pi\hbar\left(n_\vartheta + \frac{1}{2}\right), \quad n_\theta = 0, 1, 2,\ldots \quad (41)$$

where $\theta_1$ and $\theta_2$ are the classic turning points. The condition (41) results in the squared angular momentum eigenvalues [34],

$$M_l^2 = \left(l + \frac{1}{2}\right)^2 \hbar^2, \quad (42)$$

where $l = n_\theta + m$, $m = 0, \pm 1, \pm 2,\ldots$ – asymutal quantum number. Eigenvalues (42) differ from the ones, $L^2 = l(l+1)\hbar^2$, obtained from solution of the Schrödinger equation. The eigenvalues (42) are universal for all central potentials [33].

The radial equation (39) is solved by the same QC method in the complex plane. The problem has two turning points and the QC quantization condition is [33,34]

$$\oint_C \sqrt{\frac{s - m_-^2}{4s}\left[s - \left(m_+ - \frac{a(r)}{r}\right)^2\right] - \frac{M_l^2}{r^2}}\,dr = 2\pi\hbar\left(k + \frac{1}{2}\right), \quad (43)$$

where $k = 0, 1, 2,\ldots$ The phase-space integral (43) is calculated with the use of the residue theory, method of stereographic projection and the property of the asymptotic freedom $\alpha_S(r \to 0) \to 0$ [24] that gives, for the squared mass,

$$M_N^2 = \left(\sqrt{\varepsilon_N^2} + \sqrt{(\varepsilon_N^2)^*}\right)^2 \equiv \left(w_N + w_N^*\right)^2, \quad (44)$$

where $\varepsilon_N^2 = m_\|^2(1 - v_N^2) + im_\| m_\cdot v_N$, $m_\cdot = m_+/2$, $v_N = a_\infty/2N$, $a_\infty = 2/[3b_0\ln(2m_g/\Lambda)]$, $N = (k + \frac{1}{2}) + |l + \frac{1}{2}|$. In the case if $m_1 = m_2$, the (44) takes the form $M_N^2 = 4(p_N^2 + m^2)$, where $p_N = imv_N$. The formula (44) is the first asymptote for the squared system mass.

The radial equation (39) for the potential (10) is

$$\left\{\frac{d^2}{dr^2} + \frac{s - m_-^2}{4s}\left[s - \left(m_+ + V_{q\bar{q}}\right)^2\right] - \frac{M_l^2}{r^2}\right\}R(r) = 0. \quad (45)$$

The problem has four turning points and the QC quantization condition is [23]

$$\oint \sqrt{\frac{s - m_-^2}{4s}\left[M^2 - \left(m_+ + V_{q\bar{q}}\right)^2\right] - \frac{M_l^2}{r^2}}\,dr = 4\pi\hbar\left(k + \frac{1}{2}\right). \quad (46)$$

The integral (46) is calculated analogously to the above one (39) with the help of the residue theory that results in the cubic equation, for $s = M^2$,

$$s^3 + a_1 s^2 + a_2 s + a_3 = 0. \quad (47)$$

$$a_1 = 16a_\infty\sigma - m_-^2, \quad a_2 = 64\sigma^2\left(a_\infty^2 - \tilde{N}^2\right) - 16a_\infty\sigma m_-^2,$$

$$a_3 = -(8a_\infty\sigma m_-)^2, \quad \tilde{N} = N + (k + \tfrac{1}{2}).$$

The equation (47) has three (complex in general case) roots. The real part of the first root, $\mathrm{Re}\{s_1\}$, gives the physical solution,

$$M_N^2 = \mathrm{Re}\{s_{N,1}\} = \begin{cases} 2\sqrt{-p}\cos(\varphi/3) - a_1/3, & Q < 0; \\ -a_1/3 \quad (q = 0), & Q = 0; \\ f_1 + f_2 - a_1/3, & Q > 0, \end{cases} \quad (48)$$

where $\varphi = \arccos[-q/\sqrt{(-p^3)}]$, $p = -a_1^2/9 + a_2/3$, $q = a_1^3/27 - a_1 a_2/6 + a_3/2$, $Q = p^3 + q^2$, $f_1 = (-q + \sqrt{Q})^{1/3}$, $f_2 = (-q - \sqrt{Q})^{1/3}$.

The two exact asymptotic solutions (44) and (48) are used to derive the quark-antiquark mass formula. The interpolation procedure [10,35] for these two asymptotes is used to derive the mass formula that gives:

$$M_N^2 = \left(\sqrt{\varepsilon_N^2} + \sqrt{(\varepsilon_N^2)^*}\right)^2 + \mathrm{Re}\{s_{N,1}\}. \quad (49)$$

The mass formula (49) is used to describe the mass spectra of both light and heavy mesons.

# 6. Results and discussion

To demonstrate the efficiency of the model we compute the masses of the families of $\rho$, $K^{0*}$, $D^\pm$ and $B^{0*}$ Regge trajec-tories (tables 1 – 4, where masses are in MeV). The free fit to the data [1] show a good agreement for the light and heavy meson mass spectra. Note, that the gluon mass in the independent fitting of the data, $\mu_g = 416$ MeV, is the same, for all mesons and glueballs [24,25]. The $d$ quark effective mass is also practically the same, i.e., $m_d \approx 394$ MeV, for light and heavy mesons.

*Table 1. The ρ-family meson states*

| State $(n_r+1)^{2S+1}L_{Jz}$ | | $m_{theor}$ MeV/c² Formula (49) | $m_{exp}$ МэВ/c² Data [1] | Parameters |
|---|---|---|---|---|
| $1^3S_1$ | $\rho(770)$ | 775.5 | 775.26 ± 0.34 | |
| $1^3P_2$ | $a_2(1320)$ | 1313.6 | 1318.3 ± 0.6 | |
| $1^3D_3$ | $\rho_3(1690)$ | 1689.2 | 1688.8 ± 2.1 | $\alpha_\infty/2 = 1.604$ |
| $1^3F_4$ | $a_4(2040)$ | 1989.7 | 1996.3 ± 10 | $\sigma/2 = 135$ МэВ² |
| $1^3G_5$ | $a_5(1G)$ | 2247.9 | – | $m_u = 199$ МэВ |
| $2^3S_1$ | $\rho(2S)$ | 1671.4 | 1720.0 ± 20 | $m_d = 395$ МэВ |
| $3^3S_1$ | $\rho(3S)$ | 2237.5 | – | $\Lambda = 510$ МэВ |
| $4^3S_1$ | $\rho(4S)$ | 2682.2 | – | $\mu_g = 416$ МэВ |

*Table 2. The $K^{0*}$-family meson states*

| State $(n_r+1)^{2S+1}L_{Jz}$ | | $m_{theor}$ MeV/c² Formula (49) | $m_{exp}$ МэВ/c² Data [1] | Parameters |
|---|---|---|---|---|
| $1^3S_1$ | $K^*(2010)$ | 895.9 | 895.81 ± 0.3 | |
| $1^3P_1$ | $K^*(2460)$ | 1431.6 | 1431.4 ± 1.6 | |
| $1^3D_3$ | $K^*(1D)$ | 1779.3 | 1776.7 ± 7.0 | $\alpha_\infty/2 = 1.558$ |
| $1^3F_4$ | $K^*(1F)$ | 2047.1 | 2045.0 ± 9.0 | $\sigma/2 = 117$ МэВ² |
| $1^3G_5$ | $K^*(1G)$ | 2275.1 | – | $m_d = 395$ МэВ |
| $2^3S_1$ | $K^*(2S)$ | 1722.4 | 1.717 ± 9.0 | $m_s = 657$ МэВ |
| $3^3S_1$ | $K^*(3S)$ | 2241.7 | – | $\Lambda = 503$ МэВ |
| $4^3S_1$ | $K^*(4S)$ | 2643.8 | – | $m_g = 416$ МэВ |

*Table 3. The $D^\pm$-family meson states*

| State $(n_r+1)^{2S+1}L_{Jz}$ | | $m_{theor}$ MeV/c² Formula (49) | $m_{exp}$ МэВ/c² Data [1] | Parameters |
|---|---|---|---|---|
| $1^3S_1$ | $D^*(2010)$ | 2010.3 | 2010.3 ± 0.13 | |
| $1^3P_1$ | $D^*(2460)$ | 2464.4 | 2464.30 ± 1.60 | |
| $1^3D_3$ | $D^*(1D)$ | 2818.9 | – | $\alpha_\infty/2 = 1.083$ |
| $1^3F_4$ | $D^*(1F)$ | 3124.1 | – | $\sigma/2 = 221$ МэВ² |
| $1^3G_4$ | $D^*(1G)$ | 3398.6 | – | $m_d = 395$ МэВ |
| $2^3S_1$ | $D^*(2S)$ | 2789.4 | – | $m_c = 1143$ МэВ |
| $3^3S_1$ | $D^*(3S)$ | 3380.2 | – | $\Lambda = 403$ МэВ |
| $4^3S_1$ | $D^*(4S)$ | 3874.7 | – | $m_g = 416$ МэВ |

**Table 4.** The $B^{0*}$-family meson states

| State $(n_r+1)^{2S+1}L_{Jz}$ | | $m_{theor}$ MeV/c² Formula (49) | $m_{exp}$ МэВ/c² Data [1] | Parameters |
|---|---|---|---|---|
| $1^3S_1$ | $B^*(5280)$ | 5325.1 | 5325.1 ± 0.4 | |
| $1^3P_1$ | $B^*(5747)$ | 5743.0 | 5743.0 ± 5.0 | |
| $1^3D_3$ | $B^*(1D)$ | 6127.4 | – | $\alpha_\infty/2 = 0.879$ |
| $1^3F_4$ | $B^*(1F)$ | 6492.6 | – | $\sigma/2 = 604$ МэВ² |
| $1^3G_4$ | $B^*(1G)$ | 6841.7 | – | $m_d = 395$ МэВ |
| $2^3S_1$ | $B^*(2S)$ | 6101.6 | – | $m_b = 3493$ МэВ |
| $3^3S_1$ | $B^*(3S)$ | 6823.9 | – | $\Lambda = 139$ МэВ |
| $4^3S_1$ | $B^*(3S)$ | 7486.4 | – | $m_g = 416$ МэВ |

The parameter values in the potential (10), $\alpha_S$ and $\sigma$ are shown in tables 1 – 4 to be half that of the fitted values, which correspond to open strings. It was argued [36] that mesons and baryons can be described as rotating open strings in holographic backgrounds; closed strings, should be the duals of glueballs in hadron physics. A basic prediction of the closed string model is that the slope of Regge trajectories is half that of open strings. The effective tension $\sigma$ of a closed string is twice that of an open string, $\sigma_{closed} = 2\sigma_{open}$, and hence there is a major difference between the two types of strings.

Predicting the existence and shape of the Regge trajectories remains a test of the success of any phenomenological meson model. In particular, approximately-linear Regge trajectories have been shown to arise from earliest string models of mesons. Their existence is still regarded as evidence for the view of the meson as a pair of quarks connected by a gluonic string.

The RQM methods are shown to be a very powerful tool to deal with different forms of relativistic wave equations. The asymptotic method developed here provides either exact or approximate solutions, for relativistic one-particle and two-particle QC wave equations. The model supports the free particle hypothesis and those relevant to particles subject to non-trivial potentials.

## 7. Conclusion

The constituent quark picture could be questioned since potential models have serious difficulties because the potential is non-relativistic concept. However, in spite of non-relativistic phenomenological nature, the potential approach is used with success to describe mesons as bound states of quarks.

The properties of some states are still not very clear. There are theoretical indications that some of these new states could be the first manifestation of the existence of exotic hadrons (tetraquarks, molecules, hybrids etc.), which are expected to exist in QCD.

We have modeled mesons to be the bound states of two quarks interacting by the QCD-inspired funnel-type potential with the coordinate dependent strong coupling, $\alpha_S(r)$ and derived the meson interpolating mass formula (49). This approach has allowed us to describe in the unified way the centered masses of light and heavy meson states. More accurate calculations require accounting for the spin corrections, i.e., spin-spin and spin-orbit interactions. The spin-dependent corrections to the potential (10) have been considered in [10]. The mass formula (49), if considered in the complex-mass scheme [11], can be generalized to the complex eigenmasses, i.e., centered masses and total widths. These calculations can be considered elsewhere.

## References

[1] K. A. Olive et al. Particle Data Group, *Chin. Phys.*, 2014, Vol. C38, pp. 090001–090157.

[2] E. Klempt and A. Zaitsev, *Phys. Rep.* 2007, Vol. 454, pp.1–53.

[3] M. Gell-Mann, *Phys. Rev.* 1962, Vol. 125, pp. 1067–1078; *Phys. Lett.* 1964, Vol. 8, pp. 214–218.

[4] S. Okubo, *Prog. Theor. Phys.* 1962, Vol. 27, 949–953; ibid. 1962, Vol. 28, pp. 24–32.

[5] S. Nussinov, M. A. Lampert, *Phys. Rep.* 2002, Vol. 362, pp. 193–237.

[6] A. Martin, *Phys. Lett. B,* 1992, Vol. 287, pp. 251–263;

[7] D.-M. Li, B. Ma, Y.-X. Li, Q.-K. Yao, and H. Yu, *Eur. Phys. Journal*, 2004, C Vol. 37, pp. 323–337.

[8] L. Burakovsky, T. Goldman, and L P. Horwitz, *J. Phys. G*, 1998, Vol. 24, pp. 771–779.

[9] P.D.B. Collins, An Introduction to Regge Theory and High Energy Physics, Cambridge Univ. Press, Cambridge, 1977.

[10] M. N. Sergeenko, An Interpolating mass formula and Regge trajectories for light and heavy quarkonia, *Z. Phys. C*, 1994, Vol. 64, pp. 315–323.

[11] M. N. Sergeenko, Masses and widths of Resonances for the Cornell Potential, *Adv. in High Energy Phys*, 2013, Vol. 2013. Article ID 325431, pp. 1–7. {http://dx.doi.org/10.1155/2013/32543}.

[12] E. E. Salpeter, H. A. Bethe, Relativistic two body problem, *Phys. Rev.* 1951, Vol. 84, pp. 1232–1241.

[13] E. E. Salpeter, *Phys. Rev.*, 1952, Vol. 87, pp. 328–335.

[14] W. Lucha and F. F. Schoberl, Instantaneous Bethe-Salpeter Kernel for the Lightest Pseuoscalar Mesons, *Phys. Rev. D*, 2016, Vol. 93, pp. 096005–096014.

[15] E. Eichten, K. Gottfried, T. Kinoshita, J. Kogut, K. D. Lane, M. Yan, Spectrum of charmed quark–antiquark bound states, *Phys. Rev. Lett.* 1975, Vol. 34, pp. 369.

[16] G. Morpurgo, Field theory and the nonrelativistic quark model: a parametrization of meson masses *Phys. Rev. D*, 1990, Vol. 41, pp. 2865–2873.

[17] C. Semay, An upper bound for asymmetrical spinless Salpeter equations *Phys. Lett. A,* 2012, Vol. 376, pp. 2217–2221.

[18] J. Sucher, *Phys. Rev. D*, 1995, Vol. 51, pp. 5965–5973.

[19] B. Cagnac, M. D. Plimmer, L Julien and F Biraben, The hydrogen atom, a tool for metrology, *Rep. Prog. Phys.* 1994, Vol. 57(9), pp. 853–871.

[20] R. K. Bhaduri, Models of the Nucleon (From Quark to Soliton), Addison–Wesley, New York, 1988, Chap. 2.

[21] Y.-S. Huang, Schredinger-Like Relativistic Wave Equation of for the Lorentz-scalar potential, *Found. Phys.*, 2001, Vol. 31, no. 9, pp. 1287–1294.

[22] L. Durand, Behavior of relativistic wave functions near the origin for a QCD potential, *Phys. Rev.* D, 1985, Vol. 32, pp. 1257–1265.

[23] M. N. Sergeenko, Relativistic semiclassical wave equation and its solution, *Mod. Phys. Lett. A,* 1997, Vol. 12, pp. 2859–2865.

[24] M. N. Sergeenko, Glueball masses and Regge trajectories for the QCD-inspired potential. *Euro. Phys. J. C.* 2012, Vol. 72. pp. 2128–2139; {http://arxiv.org/pdf/hep-ph/1206.7099}.

[25] M. N. Sergeenko, Glueballs and the Pomeron, *Euro. Phys, Lett.*, 2010, Vol. 89, pp. 11001–11001.

[26] J. M. Cornwall, Dynamical mass generation in continuum QCD, *Phys. Rev. D,* 1982, Vol. 26, pp. 1453–1464.

[27] J. Bjorken, SLAC-PUB-6477, 1994.

[28] V. G. Kadyshevsky, *Nucl. Phys. B*, 1968, Vol. 6, pp. 125–132.

[29] F. Gross, *Phys. Rev. D*, 1974, Vol. 10, pp. 223–231.

[30] W. Lucha and F. F. Schoberl, *Int. J. Mod. Phys. A,* 1999, Vol. 14, pp. 2309–2318. arXiv:hep-ph/9812368.

[31] P. A. M. Dirac, Forms of Relativistic Dynamics, *Rev. Mod. Phys.*, 1949, Vol. 21, pp. 392–415.

[32] H. Goldstein, Classical mechanics (2nd edition Addison-Wesly Publ. Comp., Reading, Massachusetts-Menlo Park-California-London-Amsterdam-Don Mills-Ontario- Sydney, 1999).

[33] M. N. Sergeenko, Semiclassical wave equation and exactness of the WKB method, *Phys. Rev. A,* 1996, vol. 53, pp. 3798–3804.

[34] M. N. Sergeenko, Quasiclassical analysis of the three dimensional Shrodinger equation and its solution, *Mod. Phys. Lett. A,* 2000, Vol. 15, pp. 83–100.

[35] G. A. Baker, P. Graves-Morris, Pade Approximants, Gian-Carlo Rota, Editor, London, Amsterdam, Don Mills, Ontario, Sydney, Tokio: Addison-Wesly Publ. Comp., 1981, pp. 287-305.

[36] J. Sonnenschein and D. Weissman, Glueballs as rotating folded closed strings, *Journ. High Ener. Phys.*, 2015, Vol. 12, P. 011.